\documentstyle[11pt]{article}
\begin{document}
\title{VIOLATION OF BELL'S INEQUALITIES WITH A LOCAL THEORY OF PHOTONS}

\author{{\bf Patrick Suppes}\thanks{E-mail:
suppes@ockham.stanford.edu. To whom correspondence should be
addressed.} \and {\bf J. Acacio de Barros}\thanks{Permanent Address:
Physics Department, Federal University at Juiz de Fora, 36036-330 Juiz
de Fora, MG Brazil. E-mail: acacio@fisica.ufjf.br} \and {\bf Adonai
S. Sant'Anna}\thanks{Permanent Address: Mathematics Department,
Federal University at Paran\'a, C.P. 19081, 81530-900, Curitiba, PR,
Brazil. E-mail: adonai@gauss.mat.ufpr.br} \\ {\it Ventura Hall,
Stanford University,} \\ {\it Stanford, California 94305-4115}}

\date{\today}    

\maketitle

\newcounter{cms}

\setlength{\unitlength}{1mm}

\newtheorem{definition}{Definition}[section]

\begin{abstract}
We use a local theory of photons purely as particles to model the
single-photon experiment proposed by Tan, Walls, and Collett. Like Tan
et al. we are able to derive a violation of Bell's inequalities for
photon counts coincidence measurements.  Our local probabilistic theory
does not use any specific quantum mechanical calculations.

Key words: Bell's-inequalities, EPR, photons, hidden-variables,
stochastic-models. 
\end{abstract}

\section{Introduction}

	The present paper is part of a research program which concerns
a foundational analysis of phenomena usually described by quantum
electrodynamics (QED) \cite{Suppes-94a} \cite{Suppes-94b}
\cite{Suppes-96a} \cite{Suppes-96b}. Our previous papers give a
particle theory for diffraction of light and the Casimir effect. The
present paper is focused on another foundational topic. It remains to
be seen how far the program we have undertaken can be carried.

A probabilistic theory of photons with well-defined trajectories is
assumed. The wave properties come from the expectation density of the
photons. The photons are also regarded as virtual, because they are
not directly observable, including their annihilation of each other
(see assumptions bellow). What can be detected is the interaction with
matter. The meaning of {\em virtual} used here is not the same as in
QED. In summary, our basic assumptions are:

\begin{itemize}
\item{Photons are emitted by harmonically oscillating sources;}
\item{They have definite trajectories;}
\item{They have a probability of being scattered by matter;}
\item{Absorbers, like sources, are periodic;}
\item{Photons have positive and negative states ($+$-photons and
      $-$-photons) which locally interfere, when being absorbed.}
\end{itemize}

	The expected density of $\pm$-photons emitted at $t$ in the
interval $dt$ is given by
\begin{equation}
s_{\pm}(t) = \frac{A_{s}}{2}(1\pm\cos\omega t),
\end{equation}
where $\omega$ is the frequency of a harmonically oscillating source,
$A_{s}$ is a constant determined by the source, and $t$ is time. We
used $\frac{1}{2} \pm \frac{1}{2}\cos(\omega t)$ rather than
$\cos(\omega t)$, to have a density that is nonnegative for all $t$
and is between 0 and 1.  If a photon is emitted at $t'$, $0\leq t'\leq
t$, then at time $t$ the photon has traveled (with speed $c$) a
distance $r$, where
\begin{equation}
t-t' = \frac{r}{c}.
\end{equation}
The conditional space-time expectation density of $\pm$-photons for a
spherically symmetric source with given periodicity $\omega$ is:
\begin{equation}
h_{\pm} = \frac{A}{8\pi r^{2}}( 1 \pm \cos\omega( t - \frac{r}{c})),
\label{hpm}
\end{equation}
where $A$ is a real constant.

	The scalar field defined in terms of the expectation density
$h_{\pm}(t,r|\omega)$ is
\begin{equation}
{\cal E} = {\cal E}_{0}\frac{h_{+}-h_{-}}{\sqrt{h_{+} + h_{-}}},\label{field}
\end{equation}
where ${\cal E}_{0}$ is a scalar physical constant. Using (\ref{hpm}), (\ref{field}) may be rewritten for a spherically symmetric source as
\begin{equation}
{\cal E} = {\cal E}_{0}\sqrt{\frac{A}{4\pi r^{2}}}\cos\omega\left(t - \frac{r}{c}\right).\label{newfield}
\end{equation}

	Applying the standard definition of average intensity, we get the expected result
\begin{equation}
I = \left<{\cal E}^{2}\right> = \frac{{\cal E}_{0}^{2}A}{8\pi r^{2}}.
\label{intensity}
\end{equation}
Note that the standard bracket notation is used for time averaging,
i.e., taking an expectation with respect to $t$.

	Since the absorber, or photodetector, behaves periodically
with a frequency $\omega$, the probability $p_{X}$ of absorbing a
photon in detector $X$ is given by
\begin{equation}
p_{X} = \frac{C}{2}(1+\cos(\omega t + \psi)),\label{probability}
\end{equation}
where $\psi$ is an arbitrary phase that can be randomized.

	The expected number $E_t(X\pm)$ of each type of photon absorbed
by detector $X$ is the time-averaged product
\begin{equation}
E_t(X\pm) = \langle h_{\pm}^{X}(\theta)p_{X}(\psi) \rangle , 
\label{expectednumber}
\end{equation}
where $h_{\pm}^{X}(\theta)$ is the expected density of photons (with a
phase $\theta$), and $\psi$ is a phase on detector $X$. Note that
$E_t(X+)$, for example, is a random variable that is a function of
$\theta$ and $\psi$. When we take expectation with respect to the
distribution of $\theta$ we use subscripts to make clear that the
expectation is with respect to $\theta$. The averaging
is required because an absorption of an individual photon by an atom of
a photodetector takes on average several orders of magnitude longer
than the mean optical period of the photons, both theoretically and
experimentally \cite{Nussenzveig-73}.

	As we previously assumed, during the process of absorption,
photons with different states (positive and negative) annihilate each
other. So, the expected number of photons to be detected in each
detector $X$ is:
\begin{equation}
E_t(X) = |E_t(X+) - E_t(X-)|.\label{totalnumber}
\end{equation}

	We present here a violation of Bell's inequalities
\cite{Bell-64} \cite{Bell-66} \cite{Clauser-78} with a local
description of photons.

\section{Experimental Configuration}

	We are interested in the experimental setup proposed in
\cite{Tan} and also discussed in \cite{Walls}. The scheme uses two
coherent sources $\alpha_1(\theta_1)$, with phase $\theta_1$, and
$\alpha_2(\theta_2)$, with phase $\theta_2$, and a third source to be
studied, $u(\theta)$, with unknown phase.  The experimental
configuration has two homodyne detections, $(D_1,D_2)$ being one and
$(D_3,D_4)$ the other, such that the measurements are sensitive to
phase changes in $u(\theta)$.  The geometry of the setup is shown in
Figure 1.
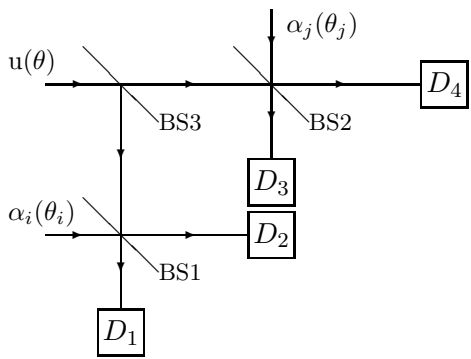
\begin{figure}[h]
\begin{picture}(100,50)
\put(40,40){\vector(1,0){5}}
\put(45,40){\vector(1,0){15}}
\put(60,40){\vector(1,0){20}}
\put(80,40){\line(1,0){10}}
\put(90,37){\framebox(6,6){$D_{4}$}}
\put(40,20){\vector(1,0){5}}
\put(45,20){\vector(1,0){15}}
\put(60,20){\line(1,0){7}}
\put(67,17){\framebox(6,6){$D_{2}$}}
\put(50,40){\vector(0,-1){10}}
\put(50,30){\vector(0,-1){15}}
\put(50,15){\line(0,-1){5}}
\put(47,4){\framebox(6,6){$D_{1}$}}
\put(70,50){\vector(0,-1){5}}
\put(70,45){\vector(0,-1){10}}
\put(70,35){\line(0,-1){5}}
\put(67,24){\framebox(6,6){$D_{3}$}}
\put(35,42){\small u($\theta$)}
\put(35,22){\small $\alpha_{i}(\theta_{i})$}
\put(72,47){\small $\alpha_{j}(\theta_{j})$}
\put(45,25){\line(1,-1){10}}
\put(45,45){\line(1,-1){10}}
\put(65,45){\line(1,-1){10}}
\put(55,34){\footnotesize BS3}
\put(75,34){\footnotesize BS2}
\put(55,14){\footnotesize BS1}
\end{picture}
\caption{Proposed experimental configuration.}
\end{figure}
In Figure 1 BS1, BS2 and BS3 are beam splitter mirrors that will
reflect 50\% of the incident photons and let 50\% of them pass. When
photons are reflected, the mirrors add a phase of $\pi/2$ to the
expected density, while no phase is added to the expected density when
photons pass through BS1, BS2 or BS3. It is easy to devise a way
to have the expected density of photons changed by a $\pi/2$ phase by
just delaying the photons that are reflected, and hence have
interacted with the mirror, by a time $T/4$, where $T$ is the period
of the photon source. We will look for correlations between the pairs
of photon detectors $(D_1,D_2)$ and $(D_3,D_4)$.

	The expected density of $\pm$-photons, generated by the
source $u(\theta)$ is:
\begin{equation}
h^u_\pm (\theta) = \frac{\beta}{2} (1 \pm \cos (\omega t + \theta)).
\label{firstudensity}
\end{equation}
The expected density coming from $u(\theta)$ at each detector is:
\begin{equation}
h_{\pm}^{D_{1}}(\theta) =
\frac{\beta}{8}\left( 1\pm\cos\left(\omega t + \theta +
\frac{\pi}{2}\right)\right),
\end{equation}
\begin{equation}
h_{\pm}^{D_{2}}(\theta) =
\frac{\beta}{8}\left( 1\pm\cos\left(\omega t + \theta + \pi\right)\right),
\end{equation}
\begin{equation}
h_{\pm}^{D_{3}}(\theta) =
\frac{\beta}{8}\left( 1\pm\cos\left(\omega t + \theta + \frac{\pi}{2}\right)\right),
\end{equation}
\begin{equation}
h_{\pm}^{D_{4}}(\theta) =
\frac{\beta}{8}\left( 1\pm\cos\left(\omega t + \theta\right)\right).\label{lasth}
\end{equation}
Note that we neglected factors of the form ${\bf k} \cdot {\bf x}$
coming from path contributions to the phase. We can do so considering
the problem completely symmetric and remembering that only phase
differences are relevant for the measurements we are proposing. 

In similar fashion, the expected density of $\pm$-photons generated by
the coherent sources $\alpha_{i}$, with phase $\theta_{i}$ and
amplitude $\alpha/2$, and $\alpha_{j}$, with phase $\theta_{j}$ and
amplitude $\alpha/2$, in each detector, is given by the following
expressions:
\begin{equation}
h_{\pm}^{D_{1}}(\theta_{i}) = \frac{\alpha}{4}\left(
1\pm\cos\left(\omega t +
\theta_{i}+\frac{\pi}{2}\right)\right),\label{firsth}
\end{equation}
\begin{equation}
h_{\pm}^{D_{2}}(\theta_{i}) = \frac{\alpha}{4}\left(
1\pm\cos\left(\omega t + \theta_{i}\right)\right),
\end{equation}
\begin{equation}
h_{\pm}^{D_{3}}(\theta_{j}) = \frac{\alpha}{4}\left(
1\pm\cos\left(\omega t + \theta_{j}\right)\right),
\end{equation}
\begin{equation}
h_{\pm}^{D_{4}}(\theta_{j}) =
\frac{\alpha}{4}\left( 1\pm\cos\left(\omega t +
\theta_{j}+\frac{\pi}{2}\right)\right),
\label{lastalpha}
\end{equation}
where we again neglected path contributions to the phase and
considered only the relevant phase at the detectors. 

	We should point out that in equations
(\ref{firstudensity})---(\ref{lastalpha}) $\alpha$ and $\beta$
are split in half at each semi-mirror, because each time a photon
reaches a mirror there is a probability of $1/2$ that the photon
passes through and a probability of $1/2$ that the photon is reflected
by the mirror.

	The probability of absorption in each detector, consistent with
equation (\ref{probability}), is given most simply by the following
equations. Some alternatives are formulated in equations
(\ref{pD1son})---(\ref{pD4nephew}) at the end of this section. 
\begin{equation}
p_{D_{1}} = \frac{C}{4}\left(2+\cos\left(\omega t + \theta_{i} +
\frac{\pi}{2}\right) + \cos\left(\omega t + \theta +
\frac{\pi}{2}\right)\right),
\label{pD1}
\end{equation}
\begin{equation}
p_{D_{2}} = \frac{C}{4}\left(2+\cos\left(\omega t + \theta_{i} + \pi
\right) +\cos\left(\omega t + \theta\right)\right),
\label{pD2}
\end{equation}
\begin{equation}
p_{D_{3}} = \frac{C}{4}\left(2+\cos\left(\omega t + \theta_{j}\right)
+\cos\left(\omega t + \theta + \frac{\pi}{2}\right)\right),
\label{pD3}
\end{equation}
\begin{equation}
p_{D_{4}} = \frac{C}{4}\left(2+\cos\left(\omega t + \theta_{j} +
\frac{\pi}{2}\right) + \cos\left(\omega t + \theta\right)\right),
\label{pD4}
\end{equation}
where $C$ is a constant that corresponds to the efficiency of the
detection process. 

	The expected number of $\pm$ photons in each detector is
given, according to equation (\ref{expectednumber}), by the following
expressions:
\begin{equation}
E_t(D_{1}^{\pm}) = \left<\left(h_{\pm}^{D_{1}}(\theta_{i}) +
h_{\pm}^{D_{1}}(\theta)\right)p_{D_{1}}\right>,
\label{ED1pm}
\end{equation}
\begin{equation}
E_t(D_{2}^{\pm}) = \left<\left(h_{\pm}^{D_{2}}(\theta_{i})
+ h_{\pm}^{D_{2}}(\theta)\right)p_{D_{2}}\right>,
\label{ED2pm}
\end{equation}
\begin{equation}
E_t(D_{3}^{\pm}) = \left<\left(h_{\pm}^{D_{3}}(\theta_{j})
+ h_{\pm}^{D_{3}}(\theta)\right)p_{D_{3}}\right>,
\label{ED3pm}
\end{equation}
\begin{equation}
E_t(D_{4}^{\pm}) = \left<\left(h_{\pm}^{D_{4}}(\theta_{j}) +
h_{\pm}^{D_{4}}(\theta)\right)p_{D_{4}}\right>.
\label{ED4pm}
\end{equation}
Equations (\ref{ED1pm})---(\ref{ED4pm}) use the fact that the
expected number of photons at a detector is simply the sum of the
number of photons from all sources. Also, in the equations above
$\left< F(t) \right> = \frac{1}{T} \int_{0}^{T} F(t) dt$, represents a
time average of the random variable $F(t)$, where $t$ is time and
$(0,T)$ is a time interval such that $\omega T\gg 1$. It is
straightforward to obtain the expressions for the total expected
number of photons in each detector from equations (\ref{hpm}),
(\ref{totalnumber}), (\ref{firsth})---(\ref{lastalpha}),
(\ref{pD1})---(\ref{pD4}), and (\ref{ED1pm})---(\ref{ED4pm}), which we
write as $I_k$, for $k=1,\ldots,4$, with $I_k$ a function of $\theta$
and $\theta_i$ or $\theta_j$:
\begin{equation}
I_1 = \left| E_t(D_{1}^{+}) - E_t(D_{1}^{-})\right| =
  \frac{C}{16}
  \left|
    \alpha
    + 
    \frac{\beta}{2}
    +
    (\alpha + \frac{1}{2}\beta)
    \cos(\theta - \theta_{i})
  \right|,
\label{firstdifference}
\end{equation}
\begin{equation}
I_2 = \left| E_t(D_{2}^{+}) - E_t(D_{2}^{-})\right| =
  \frac{C}{16}
  \left|
    \alpha
    + 
    \frac{\beta}{2}
    -
    (\alpha + \frac{1}{2}\beta)
    \cos(\theta - \theta_{i})
  \right|,
\end{equation}
\begin{equation}
I_3 = \left| E_t(D_{3}^{+}) - E_t(D_{3}^{-})\right| =
  \frac{C}{16}
  \left|
    \alpha
    + 
    \frac{\beta}{2}
    -
    (\alpha + \frac{1}{2}\beta)
    \sin(\theta - \theta_{j})
  \right|,
\end{equation}
\begin{equation}
I_4 = \left| E_t(D_{4}^{+}) - E_t(D_{4}^{-})\right| =
  \frac{C}{16}
  \left|
    \alpha
    + 
    \frac{\beta}{2}
    +
    (\alpha + \frac{1}{2}\beta)
    \sin(\theta - \theta_{j})
  \right|.
\label{lastdifference}
\end{equation}
The expressions on the right hand side of
(\ref{firstdifference})---(\ref{lastdifference}) are nonnegative,
independent of taking their absolute value, and so we subsequently
drop the absolute values.

	We are interested in the correlation between the two pairs of
detectors. First we need the variances
\begin{equation}
{\rm Var}_\theta (I_1-I_2) = E_\theta ((I_{1}-I_{2})^{2}) -
(E_\theta (I_{1} - I_{2}))^{2},
\label{var12}
\end{equation}
\begin{equation}
{\rm Var}_\theta(I_3-I_4) = E_\theta((I_{3}-I_{4})^{2}) -
(E_\theta(I_{3} - I_{4}))^{2},
\label{var43}
\end{equation}
and covariance
\begin{equation}
{\rm Cov}_\theta((I_1-I_2)(I_3-I_4)) =
E_\theta((I_{1}-I_{2})(I_{3}-I_{4})) -
E_\theta(I_{1}-I_{2}) E_\theta(I_{3} - I_{4}),
\label{covar}
\end{equation}
where $E_\theta(I_k) = \frac{1}{2\pi}\int_{0}^{2\pi}I_k d\theta$, for
$k=1,\ldots,4$, is an expectation with respect to $\theta$, with
$\theta$ uniformly distributed on $[0,2\pi]$. Thus
\begin{equation}
{\rm Var}_\theta(I_{1}-I_{2}) =
\frac{1}{512} C^{2} (\beta+2\alpha)^{2} ,
\end{equation}
\begin{equation}
{\rm Var}_\theta(I_{4}-I_{3}) =
\frac{1}{512} C^{2} (\beta+2\alpha)^{2} ,
\end{equation}
and
\begin{equation}
{\rm Cov}_\theta ((I_{1}-I_{2})(I_{3}-I_{4})) = 
-\frac{1}{512} C^{2} (\beta+2\alpha)^{2} 
\sin(\theta_{i} - \theta_{j}).
\end{equation}

	The correlation is given by
\begin{equation}
\rho_\theta (I_1-I_2, I_3-I_4) 
= \frac{{\rm Cov}_\theta ((I_{1} - I_{2})(I_{3} - I_{4}))}{\sqrt{{\rm
Var}_\theta (I_{1}-I_{2}) {\rm Var}_\theta (I_{3}-I_{4})}},
\label{corr}
\end{equation}
or
\begin{equation}
\label{viola}
\rho_\theta (I_1-I_2, I_3-I_4) = -\sin(\theta_{i} - \theta_{j}).
\label{corr2}
\end{equation}
It is easy to show that (\ref{viola}) violates Bell's inequalities
when four appropriate phases are chosen.

An examination of the derivation of (\ref{corr2}) shows that without
serious change it holds simply for a classical field as
(\ref{newfield}). Details and discussion can be found in
\cite{fieldpaper}. In the case of both Var$_\theta$ and Cov$_\theta$,
it is important to note that if we computed the correlation with
respect to $t$ rather than $\theta$, we would get different
results. It is easy to show, for example, that 
\begin{equation}
\mbox{Var}_\theta E_t(D^\pm_1)  \neq \mbox{Var}_t E_\theta(D^\pm_1) .
\end{equation}
In contrast, the order of $\theta$ and $t$ does not matter in
analyzing the data of discrete photon counts, in Section 3. 

An attentive reader may object to our expression for the probability
of detection, because we assume that the detector has the same
probability to oscillate in phase with the noncoherent source as it
has to oscillate with the coherent source, and that may bring some
non-local characteristics to the model. We can respond to this by
examining the following probability for absorption.
\begin{equation}
p_{D_{1}} = \frac{C}{2}
\left(1+\cos\left(\omega t + \theta_{i} +
\frac{\pi}{2}\right)\right),
\label{pD1son}
\end{equation}
\begin{equation}
p_{D_{2}} = \frac{C}{2}\left(1+\cos\left(\omega t +
\theta_{i}\right)\right),
\label{pD2son}
\end{equation}
\begin{equation}
p_{D_{3}} = \frac{C}{2}\left(1+\cos\left(\omega t +
\theta_{j}\right)\right),
\label{pD3son}
\end{equation}
\begin{equation}
p_{D_{4}} = \frac{C}{2}\left(1+\cos\left(\omega t + \theta_{j} +
\frac{\pi}{2}\right)\right).
\label{pD4son}
\end{equation}
The probabilities above have no term on $\theta$, but depend only on
the phase of the coherent sources. This fact have the effect of
wiping out all influences that the non-coherent source have on the
detectors, and hence putting only local parameters, like $\theta_i$ or
$\theta_j$, depending on the detector, in the probability for
detection.  If we redo the computations for the correlation with the
probabilities above, we end up with the same correlation function for
a pair of homodyne detections. In fact, to point out the robustness of
the result in face of the choice of probability for detection, we may
examine the following set of probabilities.
\begin{equation}
p_{D_{1}} = \frac{C}{2}\left(1+\frac{\alpha\cos\left(\omega t +
\theta_{i} + \frac{\pi}{2}\right) + \beta\cos\left(\omega t + \theta +
\frac{\pi}{2}\right)}{\alpha+\beta}\right),
\label{pD1nephew}
\end{equation}
\begin{equation}
p_{D_{2}} = \frac{C}{2}\left(1+\frac{\alpha\cos\left(\omega t +
\theta_{i} + \pi \right) +\beta\cos\left(\omega t +
\theta\right)}{\alpha+\beta}\right),
\label{pD2nephew}
\end{equation}
\begin{equation}
p_{D_{3}} = \frac{C}{2}\left(1+\frac{\alpha\cos\left(\omega t +
\theta_{j}\right) +\beta\cos\left(\omega t + \theta +
\frac{\pi}{2}\right)}{\alpha+\beta}\right),
\label{pD3nephew}
\end{equation}
\begin{equation}
p_{D_{4}} = \frac{C}{2}\left(1+\frac{\alpha\cos\left(\omega t +
\theta_{j} + \frac{\pi}{2}\right) + \beta\cos\left(\omega t +
\theta\right)}{\alpha+\beta}\right).
\label{pD4nephew}
\end{equation}
The above expressions would have a different physical
interpretation from the previous two presented. Each phase is given
a probability that is proportional to the amplitude of the source with
the corresponding phase. The stronger the source, the more probable to
find the detector with the same phase. It is again easy to show that
if we use these probabilities we get the same correlations as
before.

\section{Photon Counts that Violate Bell's Inequalities}

In this section we are going to use the previous result to model
discrete photon counts in such a way that they violate Bell's
inequalities.  For this, we define two new discrete random variables
$X=\pm 1$ and $Y=\pm 1$. These random variables correspond to nearly
simultaneous correlated counts at the detectors, and are defined in
the following way.
\begin{equation}
X = \left\{  \begin{array}{ll}
            +1 & \mbox{if a photon is detected at $D_1$} \\
            -1 & \mbox{if a photon is detected at $D_2$} 
            \end{array}
     \right.
\end{equation}
\begin{equation}
Y = \left\{  \begin{array}{ll}
            +1 & \mbox{if a photon is detected at $D_3$} \\
            -1 & \mbox{if a photon is detected at $D_4$.} 
            \end{array}
     \right.
\end{equation}
To compute the expectation of $X$ and $Y$ we use the stationarity of
the process and do the following. First, let us note that
\begin{equation}
I_1-I_2 = N_X \cdot P(X=1) - N_X \cdot P(X=-1),
\end{equation}
where $N_X$ is the expected total number of photons detected at $D_1$
and $D_2$ and $P(X=\pm 1)$ is the probability that the random variable
$X$ has values $\pm 1$. The same relation holds for
\begin{equation}
I_3-I_4 = N_Y \cdot P(Y=1) - N_Y \cdot P(Y=-1).
\end{equation}
To simplify we put as a symmetry condition that $N_X=N_Y=N$, i.e., the
expected number of photons hitting each homodyne detector is the
same. But we know that
\begin{equation}
I_1 + I_2 = N \cdot P(X=1) + N \cdot P(X=-1) = N,
\end{equation}
and
\begin{equation}
I_3 + I_4 = N \cdot P(X=1) + N \cdot P(X=-1) = N.
\end{equation}
Then we can conclude from equations
(\ref{firstdifference})---(\ref{lastdifference}) that
\begin{equation}
E_d(X|\theta) = \frac{I_1 - I_2}{I_1 + I_2} = \cos(\theta-\theta_i),
\end{equation}
\begin{equation}
E_d(Y|\theta) = \frac{I_3 - I_4}{I_3 + I_4} = \sin(\theta-\theta_j),
\end{equation}
where $E_d$ represents the expected value of the counting random
variable. It is clear that if $\theta$ is uniformly distributed we
have at once:
\begin{equation}
E(X) = E_\theta(E_d(X|\theta)) = 0,
\label{expectX}
\end{equation}
\begin{equation}
E(Y) = E_\theta(E_d(X|\theta)) = 0.
\label{expectY}
\end{equation}
We can now compute Cov$(X,Y)$. Note that
\begin{eqnarray}
\mbox{Cov}(X,Y) & = & E(XY) - E(X) E(Y) \nonumber \\ 
                & = & E_\theta(E_d(XY|\theta)) - 
                      E_\theta(E_d(X|\theta))
                      E_\theta(E_d(Y|\theta))
\end{eqnarray}
and so
\begin{eqnarray}
\mbox{Cov}(X,Y) & = & \frac{1}{2\pi} \int^{2\pi}_{0} E_d(XY|\theta)
 d\theta \nonumber \\ & & - \frac{1}{2\pi} \int^{2\pi}_{0}
 E_d(X|\theta) d\theta \times \frac{1}{2\pi} \int^{2\pi}_{0}
 E_d(Y|\theta)d\theta.
\end{eqnarray}
In order to compute the covariance, we also use the conditional
independence of $X$ and $Y$ given $\theta$, which is our locality
condition:
\begin{equation}
E_d(XY|\theta) = E_d(X|\theta) E_d(Y|\theta),
\end{equation}
because given $\theta$, the expectation of $X$ depends only on
$\theta_i$, and of $Y$ only on $\theta_j$. Then, it is easy to see
that
\begin{equation}
\rho (X,Y) = \mbox{Cov}(X,Y) = - \sin(\theta_i-\theta_j).
\label{finalcorrelation}
\end{equation}
The correlation equals the covariance, since $X$ and $Y$ are discrete
$\pm 1$ random variables with zero mean, as shown in (\ref{expectX})
and (\ref{expectY}), and so $\mbox{Var}(X) = \mbox{Var}(Y)  = 1.$ It
follows at once from (\ref{finalcorrelation}) that for a given set of
$\theta_i$'s and $\theta_j$'s  Bell's inequalities are violated.

\section*{Acknowledgments}

J.~A.~B. acknowledges support from the Department of Fields and
Particles (DCP) of the Brazilian Center for Physics Research
(CBPF/CNPq). A.~S.~S. acknowledges financial support from CNPq
(Brazilian Government's Support Agency).

\end{document}